# Molecular dynamics simulation of threshold displacement energy and primary damage state in Niobium

De-Ye Lin[a,b,c*], Haifeng Song[b,c], Xi Dong Hui[a*]

[a] State Key Laboratory for Advanced Metals and Materials, University of Science and Technology Beijing, Beijing 100083, PR China

[b] Institute of Applied Physics and Computational Mathematics, Fenghao East Road 2, Beijing 100094, P.R. China

[c] CAEP Software Center for High Performance Numerical Simulation, Huayuan Road 6, Beijing 100088, P.R. China

**Abstract**

In this work, a many-body potential of Nb for radiation damage simulation was developed based on embedded-atom-method (EAM), and most of the point defects of Nb can be predicted properly by this potential. By using the constructed potential, the direction-specific threshold displacement energies (TDE) and displacement cascades up to 20 keV of Nb were performed through molecular dynamics simulations. The calculated results of TDE are in good agreement with previous work for V, Mo and experimental measurements. Lowest TDE was found in <100> direction, and local minas of TDE were found in three low-index directions, which has relation: $E_d^{[100]} < E_d^{[111]} < E_d^{[110]}$. The evolution of displacement cascades, number of the created point defects, the cascade efficiency the clustering of point defects, and temperature role of these parameters at different PKA energies were systematic investigated. It is found that the cascade efficiency is low and it is can be fitted by a power function as many published work did. The fraction of clustered point defects obtained in this work is low, and only some small clusters were formed at the end of thermal spike. As the temperature increases, the productions of point defects and cascade efficiency were somewhat decreases, however, the fraction of clustered point defects decreases more obvious.

---

[*] Corresponding author: De-Ye Lin, lin_deye@iapcm.ac.cn, Xi Dong Hui: xdhui@ustb.edu.cn

# 1. Introduction

With excellent high temperature properties such as high temperature strength, good thermal conductivity and compatibility with most liquid metal coolants, the refractory metals like Nb, W, V, Mo and their alloys were used to be considered as promising materials in fusion energy, especially for fission type space reactor [1]. Several Nb alloys, such as Nb-1Zr, C-103(Nb-10Hf-1Ti), PWC-11(Nb-1Zr-0.1C), were developed and tested from early 1960's SNAP-50 program to 1990's SP-100 program [2]. However, there are serious safety and waste disposal considerations at all temperature of the Niobium alloys, and if were exposed to air, the alloys have a risk of severe oxidation at high temperatures [3]. These disadvantages limit the applications of Nb and its alloys in the nuclear industry. Nevertheless, for its low capture cross section for thermal neutrons, Nb still plays as an important alloying element in many nuclear structure materials such as zirconium alloys. For example, in alloys such as ZIRLO (Zr-1%Nb-1%Sn-0.1%Fe) [4], M5 (Zr-1%Nb-O)[5], Zr-2.5Nb[6], E110 (Zr-1%Nb) and E635 (Zr-1.3%Sn-1%Nb-0.4%Fe) [7], Nb is an important addition to improve the strength and exists as a $\beta$ phase stabilizer. Historically, most of researches on Nb and its alloys under irradiated conditions were investigated by experiment and concentrated on swelling and tensile properties[3,8]. However, the knowledge of primary radiation effects and its microstructure evolution in Niobium is still limited. As one of the powerful tools to investigate the primary damage state, molecular dynamics (MD) simulations based on valid many-body potentials have been performed to help the understanding of atomic displacement cascades in many metals and alloy systems. Brief reviews about these results would be found in References [9] and [10]. When it comes to the refractory metals, there are also some computer simulations of irradiation damage have been performed for Mo by Pasianot et al. [11], Starikov et al.[12], and Selby[13], for W by Park et al. [14], Fikar et al.[15], and Fu et al. [16], for V by Morishita et al.[17], Alonso et al. [18] and Björkas et al.[19]. Surprisingly, no work has been reported so far for Nb.

The accuracy of classical MD simulations depends entirely on the interatomic potential. Thus, to guarantee the reliability of MD simulations, a reasonable potential must be constructed in advance. Recently, there are several many-body potentials of Nb have been developed by previous researchers in the framework of Finnis-Sinclair potentials (FS potentials) [20-22], the embedded-atom-method (EAM) [23-25], the modified EAM model (MEAM) [26, 27] and analytical modified EAM model (AMEAM) [28]. However, the performance of these potentials in the radiation damage simulations has never been tested.

In our previous work, a many-body potential of Nb were also fitted by EAM model (marked by pot-1 in this work) [29]. Most static properties of Nb including cohesive energy, lattice constants, elastic constants, bulk modulus (and its first derive with pressure), vacancy properties, formation energies of several low-index surfaces can be reproduced very well. Some thermodynamics properties like thermal expansion coefficient, heat of fusion, specific heat capacity and melting point (estimated by one-phase method) can be also properly predicted through molecular dynamics simulations. However, in the following work we found that the most stable self-interstitial configuration predicted by this potential is <100> dumbbell configuration. It is obviously erroneous compared with others' models [21, 27, 28] in bcc metals. Thus, we refitted the potential based on the same model, but with some minor adjustments. The newly constructed potential is marked as pot-2 in this work. To ensure the potential be valid in high energy radiation damage simulations, the pair potential of pot-2 was corrected by a universal screened Coulomb potential proposed by Ziegler, Biersack and Littmark (*i.e.* ZBL function) [30] in the small interatomic spacing range. A four parameters exponential potential was chosen to join the ZBL function and normal pair potential. Details will be presented in Section 2. By using the newly constructed potential, most properties we calculated were comparable with pot-1. The most stable self-interstitial configuration predicted by pot-2 is <110> dumbbell, which is consistent with that by others' models [21,27,28] and results in bcc Fe [31, 32], Mo [11].

In this work, the threshold displacement energies (TDE) of Nb along different knock-on directions were calculated by MD simulations at 50 K, 300 K and 600 K. From the calculated results, we concluded that TDE of Nb has a strong anisotropic character as other bcc metals [11, 33, 34], and the temperature has slight influence on the TDE of Nb. Furthermore, the displacement cascades initiated by primary knock-on atom (PKA) with 1, 5, 10, 20 keV energies were also performed at 300 K, 600 K and 900 K. The effects of temperatures and kinetic energies of PKA on the producing of point defects, their clustering behaviors, size distributions of point defects clusters and the evolutions of displacement cascades process were investigated systematically. All MD simulations were performed using the LAMMPS [35] code developed by Sandia National Laboratory, and the open source program OVITO developed by Stukowski [36] is used to visualize the atomic configurations. Specific details of calculations will be given in the following sections.

**2. Interatomic potential and its performance in point defects calculations**

In this work, we employ the Wadley et al.'s EAM formula [37-39] to construct the potentials of Nb. By this model, both pair potential and electron density function in this model will result in a natural cutoff procedure. Unlike most EAM potentials use the standard embedding function proposed by Johnson et al. [23], Wadley et al. used a different strategy. Two simple cubic functions were used for the embedding functions near equilibrium and in lower electron density range. These could overcome some disadvantages of the standard embedding functions [37-39]. The Wadley et al.'s EAM model has been proven to be valid for bcc, fcc and hcp transition metals, potentials of various metals and alloys have already been fitted and applied in many research areas [37-39]. Though a many-body potential of Nb based on Wadley et al.'s EAM model has been constructed by us in our previous work [29], we found that it shows poor performance in the predictions of the stability of self-interstitial atom (SIA) configurations. The old potential (marked as pot-1) predicts that the most stable SIA of Nb is <100> dumbbell type, which is obviously wrong compared with results by others' models [21, 27, 28]. Thus, a new potential were refitted with minor modifications. In this work, the embedding function near equilibrium was refitted in [$0.9\rho_e$, $1.15\rho_e$] range instead of in [$0.85\rho_e$, $1.15\rho_e$] range as pot-1 did, where $\rho_e$ is the equilibrium electron density. The fitting scheme is the same as pot-1 did. Readers could refer the analytic formulas of EAM potentials in Wadley et al.'s work [37-39] and our previous work [29]. The newly fitted potential of Nb is labeled as pot-2 in this work, and all of the fitted potential parameters of present potential are list in Table 1. The cut-off radius was set as 7.0 Å.

To ensure the validity of the constructed potential in radiation damage simulations, pair potential of Nb is divided into three parts: high energy part (in short range $r_{ij} \leq r_1$), intermediate part, and normal part (the original pair potential) as most work did [40]. The interatomic interaction of atoms in high energy part is described by a universal screened

Coulomb potential proposed by Ziegler, Biersack and Littmark [30], which is called ZBL function in short and written as

$$V_{short}(r_{ij}) = \frac{Z_A Z_B e^2}{4\pi\varepsilon_0 r_{ij}} \Phi(\frac{r_{ij}}{a_s}), r_{ij} \leq r_1 \quad (1)$$

where $\varepsilon_0$ is the dielectric constant, $Z_i$ is atom number of specific elements, $e$ is electron charge number, and $a_s$ is a scaled factor defined by

$$a_s = \frac{0.8854 a_b}{(Z_A^{0.23} + Z_B^{0.23})} \quad (2)$$

where $a_b$=0.529 Å, it is the Bohr radius. And the $\Phi(x)$ is a screen function, which can be defined as

$$\Phi(x) = 0.1818 e^{-3.2x} + 0.5099 e^{-0.9423x} + 0.2802 e^{-0.4029x} + 0.02817 e^{-0.2016x} \quad (3)$$

To join the normal part and the high energy part of pair potentials soothingly, an exponential function is introduced as intermediate part of pair potential

$$V_{join}(r_{ij}) = e^{(B_0 + B_1 r_{ij} + B_2 r_{ij}^2 + B_3 r_{ij}^3)}, r_1 \leq r_{ij} < r_2 \quad (4)$$

where $r_{ij} \in [r_1, r_2)$ is the intermediate part acting range. As the parameters of equilibrium part are determined, the parameters of $B_i$ can be fitted by

$$\begin{aligned} V_{join}(r_1) &= V_{short}(r_1) \\ V_{join}'(r_1) &= V_{short}'(r_1) \\ V_{join}(r_2) &= V_{normal}(r_2) \\ V_{join}'(r_2) &= V_{normal}'(r_2) \end{aligned} \quad (5)$$

where $V'(r_i)$ is the first derivative of specific part of pair potential at $r_i$. By using this strategy, the pair potential of Nb should be continuous and smooth in the whole acting range. The fitted parameters of short-range correction are also list in table 1.

In table 2, there are several properties predicts by the constructed potential pot-2, and their comparisons with results by pot-1 [29], others' models [20, 27, 41], first-principles calculations [42-46] and experiment observations [26, 27, 47-51] are also presented. It is seen that most properties of Nb could be well reproduced by present potential. Comparing with results by pot-1, the calculated melting point, heat of fusion, thermal expansion coefficient, and heat capacity have been improved by pot-2. For the structure energy differences, results by pot-2 are somewhat smaller, and seem to be underestimated compare to those by first-principles calculations [42]. However, they are very close to those by MEAM [27], in which the angular-dependent contributions of interatomic interaction have been properly considered. Nevertheless, pot-2 ensures that the most stable structure of Nb at 0 K is bcc, which is consistent with other calculations.

As we known, point defects (vacancy and interstitial) are the original productions for a material exposing in high energy radiation conditions. Thus, it is necessary to verify the performance of constructed potential in the defects calculations. In table 3 and 4, there are some typical properties of vacancy and interstitial calculated by the constructed potential and their comparisons with others' work [27, 28, 52], respectively. As table 3 shows, vacancy properties calculated by the pot-1 and pot-2 both agree with those by MEAM [27], AMEAM [28] and experimental measurements (or first-principles calculations) very well [26, 50, 53-55]. Comparing with other potentials, the self-diffusion activation energy of vacancy calculated by pot-2 are more close to the experiment observations [53,54], but with some overestimations about the migration energies [50]. Since the activation energy was calculated by $Q_{1m} = E_{1m} + E_{1v}^{f}$ and the experimental results are come from different work. Hence, it is impossible to ensure the involved properties to agree with these experiment results at the same time. For divacancy formation energies, present potential predicts the 1st- and 2nd-nearest divacancy configurations are both stable and the calculated binding energies of them are in reasonable agreement with others' work and first-principles calculations [55].

According to Hu et al.'s work [28], there are five and four migration most possible paths for 1st-neighbor (FN) divacancy configuration and 2nd-neighbor (SN) divacancy configuration in bcc metals, respectively. To explain our results more explicitly, figure to illustrate the scheme of divacancy migration mechanism of bcc metals from Hu et al.'s work based on AMEAM model [28] is reproduced here as figure 1. The migration directions for each path are followed with arrows. Because it is lack of the experimental data, the migration energies of divacancy were also calculated by us based on an EAM potential developed by Fellinger et al. (*i.e.* EAM-DFT) [52], which had included many data obtained by first-principles calculations in the fitting of potential through force matching method. As table 3 shows, the divacancy migration energies calculated by present potential are in good agreement with those by AMEAM [28] and EAM-DFT [52]. For FN divacancy, the jumps of j2, j3 and j5 have lower energies, which indicates that the divacancy migrates along these paths are more favorable, and jumps of j1 and j4 are not possible for their obvious high migration energies. For SN divacancy, jumps of j1 and j2 have lower energies, and they are more favorable paths for SN divacancy migration, while jumps of j3 and j4 also seem to be not possible. All of these conclusions are consistent with results by AMEAM. The divacancy activation energies ($Q_{2v}$) were calculated by $Q_{2v} = E_{2v}^m + E_{2v}^f$, where $E_{2v}^f$ and $E_{2v}^m$ are the formation energy and migration energy of divacancy, respectively. As table 3 shows, the activation energies of divacancy diffusion calculated by present potential are also in good agreement with those by EAM-DFT and AMEAM. Experimentally, Mundy et al. [56] estimated the divacancy activation energy of Nb is about 5.61 eV (after unit conversion), which is also close to our calculations. Thus, the vacancy properties of Nb can be well predicted by present potential.

Interstitial is another important defects produced by radiation damage. As Johnson suggested [57], there are six possible stable configurations of self-interstitial atom (SIA) (<100>, <110>, <111> dumbbells, crowdion (C), octahedral (O), tetrahedral (T)) exist for bcc metals. In the table 4, the corresponding formation energies of these SIA calculated by present potential were presented. It should be noted that the values by MEAM were calculated by us based on Lee et al.'s model [27] except the value marked by asterisk for <110> dumbbell. For <110> dumbbell migration energy, they are all calculated by us. As table 4 shows, the calculated formation energies of different SIA are somewhat underestimated compare with those by FS [21, 58], AMEAM [28], EAM-DFT [52] and first-principles calculations [52, 59], but larger than those by MEAM. The newly constructed potential predicts that the <110> dumbbell configuration is most stable, and followed by <111> dumbbell, T, <100> dumbbells and O configurations. The energy difference between <110> dumbbell and <111> dumbbell is fairy small (0.08 eV), which is consistent with prediction by EAM-DFT. Based on our calculations, C configuration is not stable, it will transfer to <111> dumbbell configuration after relaxation. Though all of the empirical potentials including present potential predict

<110> dumbbell is most stable, however, contradictory results were obtained by first-principles calculations, for which <111> dumbbell configuration seems to be more stable. Since there are few experiment data about SIA, we could not judge which results would be more reasonable. Here, we just keep our predictions be consistent with most empirical potentials.

It is accepted that the migration energy ($E_I^m$) of SIA for most bcc metals are usually pretty small. For group V elements (V, Nb, Ta), their migration energies are usually less than 0.08 eV [50, 60]. Experiment shows the onset temperatures for SIA long-range migration occurs are lower than 3.8 K, 4~6 K and 6.3 K for V, Ta and Nb [50], respectively. That indicates that the $E_I^m$ of these three elements are pretty small and their values may be comparable. In this work, $E_I^m$ of <110> SIA was calculated by present potential via Johnson's translation-rotation mechanism [57]. The calculated value is 0.10 eV. This value is close to 0.079 eV of V calculated by Björkas et al. [19]. The value of V is determined by fitting of Arrhenius laws to the diffusivity data, and the data were prepared by using an ab-initio based potential developed by themselves. However, more results by others imply that the $E_I^m$ obtained should be somewhat overestimated. For example, Olsson [61] developed a series of EAM potentials for V and Ta, which predicts that the most stable SIA is <111> dumbbell. Based on their potentials, the calculated $E_I^m$ of <111> dumbbell are 0.016 eV and 0.022 eV for V and Ta, respectively. It is worthy noting that <111> direction is the most close-packed direction for bcc metals, thereby atoms are more easy to migrate along this direction, which will result in a lower migration energy for <111> dumbbell. Although some discrepancy was found, the low migration energy of SIA calculated by present potential still implies that a fast movement of SIA for Nb, and that is consistent with the high diffusivity of SIA for Nb from experiment observations [50].

Concluding from the performance of present potential in the point defects calculations, we believed our model is valid to apply into radiation damage simulations. Based on this potential, the threshold displacement energies of Nb in different directions and some displacement cascades events initiated by PKA with different energy of Nb were performed.

**3. Threshold displacement energy of Nb**

Threshold displacement energy (TDE, $E_d$) is a key parameter for several models to estimate the radiation damage level, such as Kinchin-Pease model and NRT formula [62]. It is usually defined as the minimum kinetic energy transferred by the PKA to a lattice atom in material and results in the formation of at least one stable Frenkel pair (pair consists of one interstitial and one vacancy).

In the TDE calculations, a box consists of 20×18×22×2= 15840 atoms in bcc lattice was used, and periodic boundary conditions (PBC) was employed in all three dimensions. All simulations were performed in NVT ensemble at 50 K, 300 K, and 600 K, respectively. Before threshold displacement energy calculations were performed, the system was firstly equilibrated at target temperature for 6 ps. Then a PKA was randomizing generated near the center area of the box with velocity along a particular direction consistent with chosen knock-on energy. All runs of TDE calculations were allowed to evolve for about 12 ps (12000 steps), though previous work show the Frenkel pairs are almost stable after 3 ps for most cases [11, 34, 62]. An uneven time step strategy was used to ensure that the atom with maximal kinetic energy should not move more than 0.05 Å in one step. Through the whole simulation processes, the system was thermalized at the given temperature by applying Berendsen thermostat method [63] to the two outer layers of atoms. This will dissipate the heat introduced by displacement events. To detect the creations of defects by atomic collisions, Wigner-Size defect analysis method [64] which has been used in many works was applied.

TDE has very obvious directional anisotropy characters for crystal materials. In this work, the direction-specific TDE is defined by $E_d(\theta,\varphi)$, where $\theta$ and $\varphi$ are the angle of PKA direction from [001] and [110] crystallographic directions, respectively. We calculated directions along border of the standard cubic stereographic triangle as follows. At first, the $E_d(\theta,\varphi)$ of bcc Nb were investigated in the range of $\theta \in [0,90°]$ and keep $\varphi = 0°$, which includes the high symmetry directions of bcc structures such as [001], [111] and [110]. Then we fixed $\theta = 90°$ and increased φ from 0° to 45°, the PKA knock-on direction return to [100]. The scheme was illustrated in figure 2. As it discussed in Nordlund et al.'s work [34], for a given PKA knock-on direction, there may have no Frenkel pair formed at high recoil energy, but at lower energy stable Frenkel pairs sometimes are produced. Thus, for each calculated directions, we search the TDE with a continuous increment in energy of 2 eV until at least one stable Frenkel pair was found in the final trajectory of calculated system. Hence the calculate accuracy of $E_d$ in this work is ±1eV. It has to pay attention that the presented values of $E_d$ along different crystal directions are averaged over four parallel runs initiated by different PKA, and the average threshold energy $E_d^{ave}$ was calculated as a direct arithmetic average of all $E_d(\theta,\varphi)$ we obtained.

The direction-specific $E_d$ calculated by present potential was displayed in figure 2, and for comparisons, results of others' work for refractory metals like V [33] and Mo [11] are also included. In the figure, two results of Mo are included, which were calculated based on potentials of EAM and embedded-defect (ED) model. ED model has included an additional term to represent the angular-dependent contributions on interatomic interactions, respectively. The knock-on direction of PKA (x-axis) displayed in figure 2 is an accumulated value from [001] to [100] crystallographic direction.

As figure 2 shows, TDE has strong anisotropy for Nb, too. Based on our calculations, there are three local minas exist along [100], [111] and [110] crystallographic directions, and they have obvious relation: $E_d^{[100]} < E_d^{[111]} < E_d^{[110]}$. Besides, three peaks of TDE are found at knock-on direction are 35°, 80° and 105°. Results obtained in the calculated paths from [001] to [110] direction in (110) plane agree with results of V very well. Comparing with Mo, some discrepancies are found. There has no reliable value of Mo along [111] direction was obtained even using larger systems. Pasianot et al. [11] argued that there has replacement collision sequence occurs along <111> direction at 0 K, and no defocusing mechanism to disturb it could be produced. However, for their another calculation of Mo based on EAM potential, it agrees with our results reasonable, although some shifts (about 5°) are found for locations of minas and peaks. That because our calculation was performed at 50 K and atoms in this system should have some atomic vibrations.

The variations of TDE along different knock-on directions could be well explained by local atom arrangements of bcc structure. We define the potential barriers formed by nearest-neighbors along PKA knock-on directions as atomic "lens", and when PKA pass through these lenses, some of the kinetic energy of PKA will transfer to them as potential energy. Thus, the more atoms interact with PKA in the moving path, the more energy needed to create a stable Frenkel pair. For PKA moves along low index crystallographic directions such as <100>, <111> and <111>, it will pass through orthogonal square lens consists of four atoms ($L_{100}$), equilateral triangle lens consists of three atoms ($L_{111}$) and rectangular lens consists of four atoms ($L_{110}$), respectively. For <100> and <110> directions, the number density of corresponding lens are $1/a^2$ and $\sqrt{2}/a^2$, where $a$ is the lattice parameter of Nb. Thus, the calculated TDE along <100> is smaller than that along <110> direction. Though the number density of $L_{111}$ is $1/\sqrt{3}a^2$, which is smaller than $L_{100}$, however, a PKA need to pass through two lenses of $L_{111}$ to knock on the next atom on lattice site, which results in more energy consuming during this process. Furthermore, for bcc metals, <100> and <111> are the favorable directions for energy transmitted by assisted focusing and pure focusing mechanisms. That results in the lower $E_d$ obtained along <100> and <111> directions than <110>. For <110> direction, since the most stable interstitial predicts by present potential is <110> dumbbell, and PKA moves in this direction will certainly help the formation of <110> dumbbell. That may be a reason of a local mina of TDE formed around <110> direction by present potential.

When it comes to high index directions, the number density of atomic plane seems to less reliable in the predictions of direction-specific TDE [33], but the formation of peaks in figure 2 still can be explained with local geometry of bcc structure. In figure 4, two projections of atoms in ($\bar{1}10$) plane and (001) plane were shown. The direction-specific velocity of PKA is defined as $v(\theta,\varphi)$ similarly to $E_d$. As figure 4(a) shows, when $\varphi=0°$ and $\theta=0°$, which is [001] direction and has lowest TDE. With $\theta$ increases from 0° to 25.24°, TDE increases. That because the moving path of PKA has left from [001] direction to high index direction, thus the defocusing phenomenon is strengthened. In the range of $v(25.24\sim35.26°,0°)$, PKA has to pass through a long distance to knock on the next atoms on lattice site, more energy is sacrificed during the process. Therefore, a peak of TDE is found around this range (~35° in our calculations at 50 K), and the minor shift comes from thermal vibrations of atoms. The same processes occur for PKA moving along $v(76.74°, 0°)$ and $v(90°, 18.43°)$, which results in the formations of other two peaks in figure 3 at accumulated knock-on angle equal 80° and 105°.

Some results of TDE calculated along low index directions and its comparisons with different experiment data are summarized in table 6. In the table, the $E_d^{min}$, $E_d^{med}$ and $E_d^{ave}$ are the lowest value, median value and average value of TDE obtained from all direction-specific $E_d(\theta,\varphi)$ calculations. As it is shown, $E_d^{min}$ calculated by present potential is in reasonable agreement with experiment observations from both polycrystalline and single crystalline samples [50], the discrepancies are modest. However, the average TDE calculated by present potential just about half the experimental data, which is 78(87) eV [50]. Here, the experimental values were derived from single crystal and polycrystalline (in parentheses) damage rate measurements, respectively. We checked results from others work on refractory metals, and similar tendencies were found in many calculations by many body potentials such as FS and EAM. For V, results by Björkas et al. [19] and Depeda-Ruiz et al. [33] are 55 eV and 51 eV, and the experimental value is 88 eV; For W, result by Björkas et al. [19] is 85.4 eV and experiment value is 193 eV; For Mo, results by Pasianot et al. [11] based on potentials of EAM and ED are 43 eV and 50 eV, while experiment value is 75(82) eV. As it is shown, all of the mentioned potentials seem to underestimate the average TDE much, thus result calculated by present potential is consistent with that by these potentials. It has to pay attention that most average TDE calculated by the above potentials were always compared with the recommend values from ASTM standard [65], which is much lower than experiment data from Ref. [50] for most refractory metals. As table 6 shows, our result still seems to underestimated the TDE of Nb about 1/3. However, it's noted that the value of Nb in ASTM standard was directly set equal to that for Mo. Experimentally, the TDE of Mo is 34.5±0.5 eV at 70 K, which is obviously larger than 28.5 eV of Nb at 50 K [50]. In this work, the ratio of $E_d^{[111]}/E_d^{[100]}$ calculated by present potential at 50 K is 1.28. It falls exactly in the range of 1.0~1.29 for most bcc metals by experimental observations [50]. By comparing with experimental data and work for other refractory metals, we believe that present potential could also reproduce the threshold displacement energy of Nb properly, though some discrepancies are found and possibly further improvement of the potential is required.

To investigate the temperature dependence of TDE, MD simulations were also performed at 300 K and 600 K. Results compared with 50 K is displayed in figure 5. It can see that the temperature has little influence on TDE for Nb, except that some decrease are found in <100> direction. Qualitative results are summarized in table 6. As the table shows, with the temperature increasing from 50 K to 600 K, $E_d^{min}$ falls from 25.0 eV to 21.1 eV, which is comparable with the experiment data obtained by electron irradiation at different temperatures from transmission electron microscopy on loops studied on polycrystalline samples [50].

**4. Displacement cascades simulations**

Most of the conditions in displacement cascades simulations were set as threshold displacement energy calculations did, except that the cascades were initiated by a PKA knock-on along [135] direction. Many publications have proved that result calculated along [135] direction is representative of the average defect production for bcc metals [10, 31]. The box size and simulated time at each PKA kinetic energy level are also list in table 5. To obtain statistically meaningful results 5~10 parallel cascade events were simulated per energy. Variability of cascades was controlled by different PKAs. To obtain average results dependence on simulated time from different cascades under the same conditions, a different strategy of timestep setting was applied during the displacement cascades simulations. At first, we performed a displacement cascades at 20 keV by an uneven timestep scheme as TDE calculations did. We carefully checked the variations of timestep for every 100 steps, and found that when the timestep was set as 0.01, 0.1 and 1 fs in the range of 1~10000 steps, 10001~20000 steps and 20001~ after the PKA was generated, atoms in the system will not move more than 0.05 Å in one step. Thus, this timestep scheme was employed in all displacement cascades simulations of this work. To compare with previous work for V and W [19], interstitial closer than third nearest neighbor (3NN) distance and vacancy closer than second nearest neighbor (2NN) distance were counted in the same defect cluster.

In figure 5, the variations of averaged number of Frenkel pairs ($N_{FP}$) created by PKA at different energies with simulation time were displayed. As it shows, more defects are created with higher PKA knock-on energy and the displacement spikes occur at about 0.3~0.8 ps depend on PKA energy. With the simulate time increased, most of the defects quickly varnished by recombination in no more than 3~10 ps, and only a few radiation-induced interstitials and vacancies can survive after ~20 ps quenching process. These are consistent with much work on molecular dynamics simulations for metals [1, 16, 19, 40]. The distributions and evolutions of defects created by different energies are displayed as figure 6. We can found that a great deal of the Frenkel pairs was produced at peak times, and vacancies are more likely to be created near the center of the cascades, however, interstitials tends to distribute around the outer regions of the displacement spikes. The shape of cascade structure seems to have correlations with the final state of defects. According to the study by Becquart et al. [66] on the influence of interatomic potentials for iron cascades, they found that the shape of cascade is sensitive with potentials. If the cascade structure at peak state is dens and isotropic, less number of stable point defects will be created at the end of thermal spike. As figure 6 shows, both cascade structures at peak state created by 5 keV and 20 keV PKA are dense and isotropic based on our calculations. Thus, fewer defects would be produced as figure 6 shows.

Besides, we compared our results of defects produced by 10 keV PKA with Björkas et al.'s work for V and W [19], we found that the dependence of defects and simulated time obtained for Nb is similar to that for W, except that the number of defects of Nb is about twice than W. The results of V seem to be dissimilar with Nb and W, but comparable number of defects was obtained in the final state as Nb did. Based on present potential and potentials by Björkas et al., the vacancy and interstitial formation energies for Nb are 2.76 eV and 3.03-3.60 eV, for V are 2.51 eV and 3.3-4.2 eV, whereas for W these values are 3.56 eV and 9.548-11.68 eV, respectively. W has higher formation energy while Nb and V have comparable formation energy. The higher values these values are, the fewer point defects will be produced under the same PKA energy. That should explain the less number of residual defects were created for W and comparable results were obtained for Nb and V based on the above potentials.

It has been proved that $N_{FP}$ has a simple power function with cascade energy ($E_{MD}$) as

$$N_{FP} = A(E_{MD})^m \tag{6}$$

By least-square method, results of Nb calculated from 1 keV to 20 keV were fitted and shown in figure 7. As it shows, the results of $N_{FP}$ calculated by present potential for Nb can be well fitted by the power function. The fitted values of *A* and *m* are 3.645 and 0.767, 3.670 and 0.690, 4.218 and 0.649 at 300 K, 600 K and 900 K, respectively. The differences of fitted lines obtained at different temperatures are negligible; however, somewhat fewer defects were produced at higher temperature, which is in accordance with many MD simulations for Fe [66, 67], Zr [68], W [16]. That is because the mobility of vacancy is elevated at high temperature, which will increase its recombination with interstitials or self-clustering behavior. These effects will outrun the effects of temperature on the displacement process that leads to more productions of Frenkel pairs [50].

Typically, the ratio of $N_{\text{FP}}/N_{\text{NRT}}$ is always referred as cascade efficiency, where $N_{\text{NRT}}$ is the defects number estimated by NRT formula [69]. NRT formula has been used in many work to estimate the damage level of materials under radiation condition, and it is written as: $N_{NRT} = 0.8\,E_{\text{MD}}/E_{\text{d}}^{*}$, where $E_{\text{d}}^{*}$ is the effective threshold displacement energy. In this work, we took the value as 60 eV recommended by ASTM standard [67]. The derived cascade efficiency averaged over our calculations for different energies is displayed in figure 8(a). As figure 8(a) shows, the cascade efficiency is decreased as temperature increase, and comparable results were obtained for 600 K and 900 K. In figure 8(b), we also compared our results of Nb at 300 K with other work for refractory metals. It is seen that the cascade efficiency is very sensitive to interatomic potentials, different results were obtained by Pasianot et al. [11] and Selby et al. [13] for Mo. Nevertheless, results obtained by these calculations seem to have similar tendency, the cascade efficiency seems to be decreased with energy increased.

The in-cascade clustering of defects by radiation damage has strongly influence on the microstructure evolution of materials, and further influences their mechanic performance under radiation conditions. Thus, the clustering behavior of interstitials and vacancies of Nb were also investigated in this work. In figure 9, the fraction of point defects forming clusters at the final state of Nb is shown. Values displayed in the figure were averaged value over the last 1000 quenching steps. As it shows, both results of vacancy and interstitial are more scattered than the obtained number of Frenkel pairs in figure 7. That because the number of stable Frenkel pairs could be produced at given energy is linear with PKA energy, which as NRT formula indicated, but the arrangement of point defects can be in many different ways [1]. Nevertheless, some trends of in-cascades clustering of point defects still can be extracted. As figure 9 shows, the clustering of vacancy seems to be really stable as PKA energy increased, though a bit of decreasing was found. These are in reasonable agreement with several simulations on refractory metals such as V by Björkas et al. [19], and Mo by Selby et al. [13]. According to Björkas et al.'s point of view [19], for refractory metals such as Nb, which has a high melting point (2650 K predicted by present potential), the recrystallization front formed by thermal spike will move faster, and that will lead to a low vacancy cluster fraction. The melting point of Nb is close to V (2350 K predicted by Björkas et al.'s potential), therefore, only slightly decrease of fraction of clustered vacancy was found as V did by Björkas et al.. In contrast with vacancy, the clustering of interstitial seems to be more favorable at higher PKA energy than at lower PKA energy, but the fractions of clustered interstitial are still low. According to Stoller said in Reference [1], the formation of SIA clusters are determined well before the onset of the thermal spike phase. For displacement cascades simulated by present potential, most of thermal spike phases seem to have dens and isotropic shape and few sub-cascades were formed even at 20 keV PKA energy, which results in the less possibility of the interstitial clustering. Typical structure of thermal spike phases at peak time of displacement cascades are shown in figure 6.

    The size distribution of point defect clusters is shown in figure 10. Cluster size is defined as the number of defects in the cluster, and the data displayed in the figure were obtained by averaging over all simulations at each PKA energy. As it shows, no interstitial cluster larger than 6 was found even at high temperature, only limited small cluster were formed in the final state of primary damage simulations (>20 ps). However, clusters larger than 10 vacancies were found in the simulations of 20 keV PKA energy at different temperatures. At 600 K, the frequency to find vacancies clusters larger than 10 is higher than at 300 K and 900K, but at 900 K, clusters with more than 10 vacancies was found even in one of the simulation of 10 keV PKA energy. Thus, the temperature dependence of interstitial clustering seems to be obscure. Since the data of point defects clustering are scattered as mentioned above, more simulations need to perform for further investigations. Nevertheless, the cluster size of point defects of Nb obtained in this work is fairy small, which predicts the radiation-induced

growth, swelling and formation of dislocation loops should be limited based on our calculations.

Comparing with results at different temperatures, we found that the point clustering decreases as the temperatures increases as figure 9 shows. The dependence of vacancy clustering with temperature obtained by us for Nb is consistent with conclusion obtained by Stoller for Fe with 2NN criterion applied in cluster determination [1]. Whereas, for interstitial cluster, most results by others [67,68] seems to be contradict with us. But it has to be noted that most of conclusions were obtained from simulations of Fe [67] and Zr [68], which have different crystal structures or dissimilar physical properties with refractory metals like Nb. For MD simulations of Fe [67] and Zr [68], high frequency of large point defects clusters were found at the end of thermal spike. These clusters are sessile and do not tend to dissociate or transform into glissile clusters even at higher temperature. In contrast, they act as sinks for interstitials and vacancies. As temperature increases, the gatherings of point defects become more severe. Thus, fractions of clustered point clusters increases as temperature increases. In this work, the point clusters of Nb obtained by us are small, which seem to be mobile and glissile [70]. Although the vacancy clusters are almost immobile at low temperature, they become mobile at high temperature (600 K and 900 K in this work). These vacancy clusters will recombine with interstitial clusters, which results in a decrease of point defects cluster. Therefore, it is expected that the fraction of clustering point defects will drop at higher temperature.

Experimentally, the concentration of radiation-induced point defects in materials can be measured by electronic resistivity changes observations. Based on experiment observations [50], the resistivity changes of Nb is very small (5.6 n$\Omega$·m at 24 K and 7.4 n$\Omega$·m at 19 K by neutron irradiation), which is comparable with Mo (4~9.5 n$\Omega$·m), W (5.6~9.2 n$\Omega$·m) and Ta (4.6~6.9 n$\Omega$·m). That implies the high radiation damage resistance of Nb. Based on our calculations, the number of stable Frenkel pairs are low, which also predicts fairy well radiation damage resistance of Nb. Thus, our predictions of cascade efficiency by present potential are consistent with experimental measurements.

In Jones et al.'s work [71], the damage of Nb induced by 14-MeV neutrons had been evaluated by transmission electron microscopy. They found that the microstructure of radiated

Nb consists of a fairly uniform distribution of small defects clusters. That agrees with our calculations which predicted a similar tendency. The clustering of vacancies results in void formation and swelling of radiated materials. Experimentally, it is found that the swelling of radiated Nb is limited [1, 70] at $T=0.2\sim0.5T_\mathrm{m}$, where $T_\mathrm{m}$ is the melting temperature. Based on our calculations, both the cluster size and fraction are small even at 900 K. That is also in qualitative agreement with the experiment observations.

The direct experimental observations of temperature dependence of point defects clustering of Nb is lacking, however, in Singh et al.'s review [70], experiment data on temperature dependence of cluster density in neutron irradiated Mo was presented. Singh et al. found that a rapid decrease in cluster density of Mo with increasing irradiation temperature and the cluster density of Mo is low. As we mentioned above, the radiation damage level of Mo is expected to be comparable with Nb, and their melting point is close. Therefore, we believe that the temperature dependence of point defects clustering predicted by present potential for Nb is consistent with experimental measurement.

## 5. Summary

In this work, a many body potential of Nb (present potential) was reconstructed within the framework of EAM. To use this potential in radiation damage simulation, pair potential of the model was corrected the ZBL screening function in the short range spacing. Comparing with others' models, first-principles calculations and experiment data, we believe that the constructed potential could reproduce many properties of Nb reasonable. Besides, the properties of vacancy and interstitial were also calculated. Our results show that the vacancy (divacancy) formation energy, vacancy (divacancy) migration energy and action energy calculated by present potential agree with those by MEAM, AMEAM, EAM and experimental measurements. For the formation of SIA, the newly constructed potential can predict reasonable SIA formation tendency compare to our previous constructed potential (pot-1). The newly constructed potential predicts the most stable configuration of SIA is <110> dumbbell, and followed by <111> dumbbell, T, <100> dumbbells and O configurations. The most stable configuration of SIA predicted by present potential is consistent with results by most empirical many body potentials, but has some discrepancy with results by first-principles calculations, which predict <111> is the most stable SIA configuration for Nb. The migration energy of <110> dumbbell was calculated via Johnson's translation-rotation mechanism. The calculated value is 0.10 eV. This low value can imply the high diffusivity of SIA for Nb, however, it still seems to be somewhat overestimated compared with data by MD simulation and experimental observations of other refractory metals.

Based on present potential, TDE of Nb was calculated along border of the standard cubic stereographic triangle. Based on our calculations, there are three local minas exist along [100], [111] and [110] crystallographic directions, and they have relation: $E_d^{[100]} < E_d^{[111]} < E_d^{[110]}$. Besides, three peaks of TDE are found at knock-on direction are 35°, 80° and 105°. That is in good agreement with Depeda-Ruiz et al.'s calculation for V [33] and Pasianot et al.'s calculation for Mo [11]. The ratio of $E_d^{[111]}/E_d^{[100]}$ calculated by present potential at 50 K is 1.28. It falls exactly in the range of 1.0~1.29 for most bcc metals by experimental observations [50]. The formations of minas and peaks formed in the direction-specific TDE were well explained by local atomic arrangement of Nb and the number density of atomic lens which PKA pass though. Besides, the temperature dependence of TDE seems to be weak based on our calculations, but the lowest value and average value of TDE decrease as temperature increases. That is also consistent with experimental measurements very well. However, average values of TDE calculated by present potential is only about half of experiment measurements. By comparing with others' work for V [19], W [19] and Mo [11], we found that the underestimations of average TDE may be a common drawback of many-body potentials.

By molecular dynamics simulations, the displacement cascades of Nb initiated by 1 keV, 5keV, 10keV and 20 keV PKA were studied at 300 K, 600 K and 900 K. The evolution of displacement cascades, number of the created point defects, the cascade efficiency the clustering of point defects, and temperature role of these parameters at different PKA energies were systematic investigated. Based on our calculations in this work, less number of stable point defects will be created compared with NRT formulas. The cascade efficiency is also followed with pow law function as many published work did. As the temperature increases, the creations of point defects as well as cascade efficiency somewhat decreases. The fraction of clustered point defects obtained in this work is low, and only some small clusters were formed at the end of thermal spike. As temperature increases, the fraction of clustered point defects decreases. Those because the small point clusters of Nb obtained by us are mobile and glissile [70]. The vacancy clusters will become mobile at high temperature and recombine with interstitial clusters, which results in a decrease of point defects clustering. Low frequencies of large clusters were found in our calculations. No interstitial cluster larger than 6 was found, and only a few vacancy clusters larger than 10 were found. Besides, the size distribution of vacancy clusters seems to have little correlation with temperature according to our calculations; however, the temperature dependence of interstitial clustering seems to be obscure.

We also compared our results with experimental measurements of Nb and some other refractory metals. It is seen that most of our results are consistent with experiment reasonably. In conclusion, though some deviations were found, our work still predicts the good radiation resistance of Nb, we believe this work is still helpful in the understanding of the primary radiation damage of Nb.

## Acknowledgements

This work was jointly supported by the XXX.

Table 1. Values of the potential parameters of Nb used in this work.

| $m$ | $n$ | $r_e$ | $f_e$ | $\rho_e$ | $\rho_s$ |
|---|---|---|---|---|---|
| 20 | 20 | 2.858230 | 2.889832 | 30.560600 | 30.560600 |
| $A$ | $B$ | $\alpha$ | $\beta$ | $\kappa$ | $\lambda$ |
| 0.547673 | 0.851301 | 7.272435 | 4.500000 | 0.094614 | 0.363743 |
| $\rho_n/\rho_e$ | $\rho_o/\rho_e$ | $\eta$ | $F_e$ | $Fn_0$ | $Fn_1$ |
| 0.90 | 1.15 | 0.799332 | 4.954020 | -4.937144 | -0.322120 |
| $Fn_2$ | $Fn_3$ | $F_0$ | $F_1$ | $F_2$ | $F_3$ |
| 1.565574 | -3.049441 | -4.954560 | 0 | 1.647300 | -0.948727 |
| $B_0$ | $B_1$ | $B_2$ | $B_4$ | $r_1$ | $r_2$ |
| 27.709940 | -41.404510 | 25.030018 | -5.460230 | 1.38 | 2.31 |

Table 2 Some properties of Nb calculated by the constructed potential and their comparisons with others' work. In the table, $E_c$ is cohesive energy (eV), $a_0$ is lattice constant (Å), $C_{ij}$s are elastic constants (GPa), $B_0$ is bulk modulus (GPa), $\partial B/\partial p$ is the first derive of bulk modulus by pressure at 0 K, $E_{1v}^f$ is mono-vacancy formation energy (eV), $\Delta E$s are the structural energy differences (eV), $E_{surf}$ are surface formation energies (mJ/m$^2$), $T_m$ is the melting point (K), $\Delta H_f$ is the heat of fusion (kJ/mol), $C_p$ is the heat capacity (J/mol/K, 273-373 K), and $\alpha$ is the linear thermal expansion coefficient (K$^{-1}$).

| | FS [20] | TB-SMA [41] | MEAM [27] | First-principles calculations | pot-1 [29] | pot-2 (present potential) | Experiment |
|---|---|---|---|---|---|---|---|
| $E_c$ | 7.57 | 7.57 | 7.47 | | 7.57 | 7.57 | 7.57 [47] |
| $a_0$ | 3.3008 | 3.30 | 3.3024 | 3.322[42] | 3.3000 | 3.3001 | 3.3004 [48] |
| $C_{11}$ | 246.6 | 247 | 252.7 | 247.2[43] | 262.2 | 246.9 | 246.5 [48] |
| $C_{12}$ | 133.2 | 135 | 133.1 | 140.0[43] | 124.6 | 134.7 | 134.5 [48] |
| $C_{44}$ | 28.1 | 29.0 | 31.9 | 14.2[43] | 35.95 | 28.88 | 28.73[48] |
| $B_0$ | 171.0 | 172 | 173.0 | 172.3[43] | 170.2 | 172.1 | 170.2[48] |
| $\partial B/\partial p$ | | | 4.23 | 3.89[43] | 4.10 | 4.10 | 4.10, 4.72 [49] |
| $E_{1v}^f$ (eV) | 2.64 | | 2.75 | 2.92[44] | 2.75 | 2.76 | 2.75[26] |
| | | | | 2.32[45] | | | 2.7[50] |
| | | | | 2.79-2.88[46] | | | |
| $\Delta E_{bcc \to ideal\ hcp}$ | | 2.84 | 2.947 | 2.880[42] | 2.937 | 3.002 | |
| | | 0.21 | 0.164 | 0.2915[42] | 0.240 | 0.173 | |
| $\Delta E_{bcc \to fcc}$ | | 4.04 | 4.169 | 4.230[42] | 4.154 | 4.246 | |
| | | 0.22 | 0.176 | 0.3234[42] | 0.240 | 0.173 | |
| $\Delta E_{bcc \to sc}$ | | | 2.606 | | 2.568 | 2.558 | |
| | | | 0.90 | | 0.540 | 0.728 | |
| $\Delta E_{bcc \to diamond}$ | | | 5.425 | | 5.415 | 5.451 | |
| | | | 1.44 | | 2.050 | 2.158 | |
| $E_{surf}$(100) | 2046 | 1792 | 2490 | | 1893 | 1869 | |
| $E_{surf}$(110) | | 2101 | 2715 | | 2285 | 2310 | |
| $E_{surf}$(111) | | 2343 | 2923 | | 2628 | 2607 | |
| $E_{surf}$(average) | | 2079 | 2709 | | 2269 | 2262 | 2300[27] |
| | | | | | | | 2983[51] |
| $T_m$ | | 3000 | 1900 | | 2750 | 3150 | 2750[48] |
| | | | | | (2380) | (2650) | |
| $\Delta H_f$ | | | 13.5 | | 19.0 | 22.22 | 30.0[48] |
| $C_p$ | | | 26.1 | | 25.3 | 23.1 | 24.9[48] |
| $\alpha$ (273-373 K) | | | 6.4 | | 6.9 | 7.4 | 7.3[48] |

Table 3 Vacancy properties of Nb (units: eV ) calculated based on different models and their comparisons with experiment or first-principles calculations (in *italic*). For the results of MEAM and EAM, values with * are referred from their original work [Fellinger,], and others are calculated by us based on the same models.

| Nb | MEAM [27] | EAM-DFT [52] | AMEAM [28] | pot-2 (present potential) | Experiment |
|---|---|---|---|---|---|
| $E_{1v}^{f}$ | 2.75* | 3.10* | 2.76 | 2.76 | 2.75[26] |
|  |  | 3.20 |  |  | 2.7[50] |
| $E_{1m}$ | 0.57* | 0.77* | 0.64 | 0.76 | 0.55 [50] |
|  |  | 0.70 |  |  |  |
| $Q_{1m}$ | 3.32* | 3.87* | 3.40 | 3.52 | 3.6[53] |
|  |  | 3.90 |  |  | 3.7[54] |
| $E_{2v}^{f}$ (FN) |  | 6.00 | 5.16 | 5.20 | *5.62*[55] |
| $E_{2v}^{b}$ (FN) |  | 0.40 | 0.36 | 0.32 | *0.36*[55] |
| $E_{2v}^{f}$ (SN) |  | 6.13 | 5.20 | 5.27 | *5.57*[55] |
| $E_{2v}^{b}$ (SN) |  | 0.27 | 0.32 | 0.25 | *0.41*[55] |
| $E_{j1}^{FN}$ |  | 1.95 | 1.66 | 1.53 |  |
| $E_{j2}^{FN}$ |  | 0.66 | 0.63 | 0.73 |  |
| $E_{j3}^{FN}$ |  | 0.84 | 0.78 | 0.89 |  |
| $E_{j4}^{FN}$ |  | 2.72 | 1.75 | 1.86 |  |
| $E_{j5}^{FN}$ |  | 0.70 | 0.64 | 0.76 |  |
| $Q_{2v}^{FN}$ |  | 6.66-8.72 | 5.79-6.94 | 5.84-6.94 |  |
| $E_{j1}^{SN}$ |  | 0.61 | 0.61 | 0.80 |  |
| $E_{j2}^{SN}$ |  | 0.70 | 0.64 | 0.76 |  |
| $E_{j3}^{SN}$ |  | 2.77 | 2.09 | 1.93 |  |
| $E_{j4}^{SN}$ |  | 2.71 | 1.98 | 1.84 |  |
| $Q_{2v}^{SN}$ |  | 6.74-8.90 | 5.81-7.29 | 5.91-7.08 | 5.61 [56] |

Table 4 The calculated Self-interstitial formation energies (eV) of Nb in this work. By our constructed potentials, the C interstitial will transfer to be <111> dumbbell configuration after relaxation. For MEAM, values with * are refereed from Ref. [27], and others are calculated by us based on the same model. For <110> dumbbell migration energy, they are all calculated by us.

| SIA | pot-1 | pot-2 (present potential) | FS [21] | FS [58] | MEAM [27] | AMEAM [28] | EAM-DFT [52] | First-principles calculation (PBE) | |
|---|---|---|---|---|---|---|---|---|---|
| | | | | | | | | [52] | [59] |
| <100> | 2.77 | 3.54 | 4.821 | 4.85 | 3.80 | 4.44 | 4.50 | 4.76 | 5.949 |
| <110> | 3.07 | 3.03 | 4.485 | 4.54 | 2.56* 2.56 | 4.39 | 3.83 | 4.31 | 5.597 |
| <111> | 3.00 | 3.11 | 4.795 | 4.88 | 3.09 | 4.74 | 4.09 | 3.95 | 5.253 |
| O | 2.90 | 3.60 | | 4.91 | 2.56 <110> | 4.43 | 4.36 | 4.89 | 6.060 |
| T | 3.06 | 3.40 | | 4.95 | 2.56 <110> | 4.73 | 4.37 | 4.56 | 5.758 |
| C | 3.01 <111> | 3.11 <111> | 4.857 | 4.95 | 3.12 | 4.93 | 4.02 | 3.99 | |
| $E_I^m$ <110> | | 0.10 | | | 0.33 | | 0.16 | | |

Table 5 Typical simulation cell sizes used in this work.

| $E_{\text{PKA}}$ in MD (keV) | Box Size | Number of atoms | Number of Cascades | Annealing time (ps) |
| --- | --- | --- | --- | --- |
| 1 | 30×32×34 | 65280 | 10 | 20 |
| 5 | 40×42×44 | 147840 | 10 | 20 |
| 10 | 50×52×54 | 280800 | 5 | 20 |
| 15 | 68×70×72 | 685440 | 5 | 30 |
| 20 | 84×86×88 | 1271424 | 5 | 30 |

Table 6. The calculated threshold displacement energies along some typical crystallographic directions and their comparisons with experimental data. For $E_d^{min}$ and $E_d^{ave}$, the first line are results calculated by present potential, and the second and third lines are experiment data obtained under electron irradiation at different temperatures from transmission electron microscopy on loops studied on polycrystalline and single crystalline samples(*), respectively [50]; the forth line are values recommended by ASTM E521-96(2009) standard [67].

| T (K) | <100> | <110> | <111> | $E_d^{min}$ (eV) | $E_d^{med}$ (eV) | $E_d^{ave}$ (eV) |
|---|---|---|---|---|---|---|
| 50 | 25.0 | 48.1 | 34.0 | 25.0 | 35.8 | 41.1 |
|  |  |  |  | 28 |  |  |
| 300 | 22.8 | 53.6 | 35.6 | 22.8 | 36.9 | 41.1 |
|  |  |  |  | 24, 24.5 |  | 87 |
|  |  |  |  | *24 |  | *78 |
|  |  |  |  | 36 |  | 60 |
| 600 | 21.1 | 49.8 | 34.8 | 21.1 | 39.1 | 39.4 |

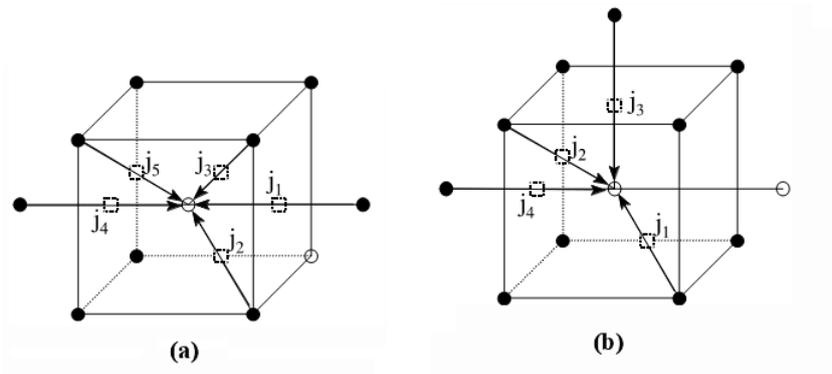

Figure 1. Possible migration paths of divacancy diffusion for Nb. The figure is reproduced from Ref. [28].
(a) FN divacancy; (b) SN divacancy

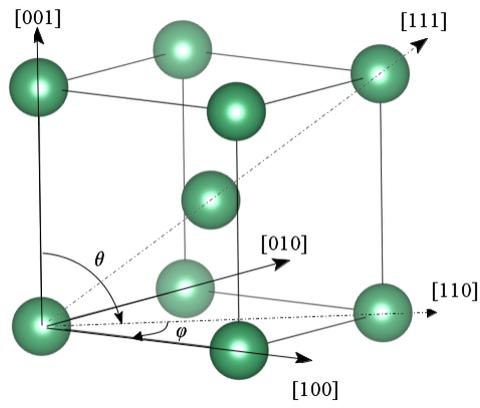

Figure 2. The direction-specific threshold displacement energy calculations scheme of Nb.

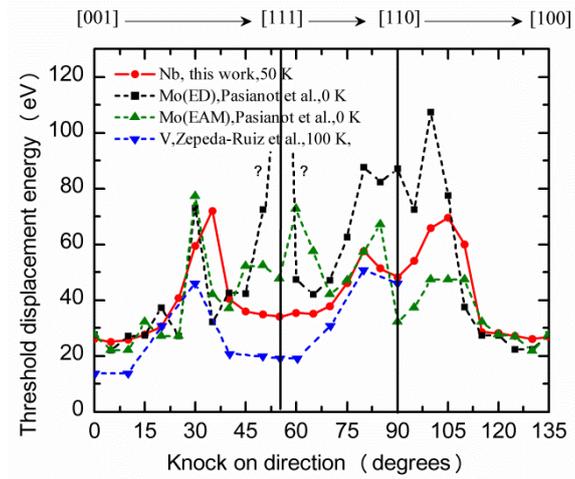

Figure 3. Dependence of the threshold displacement energy calculated by present potential for Nb along different PKA knock-on directions, and its comparisons with others' work for V [19] and Mo [11].

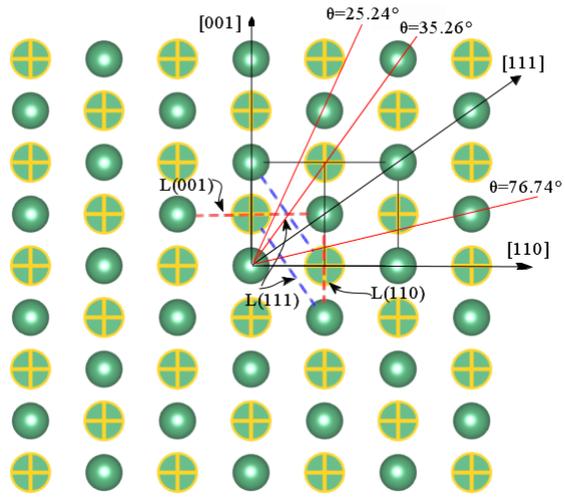

(a)

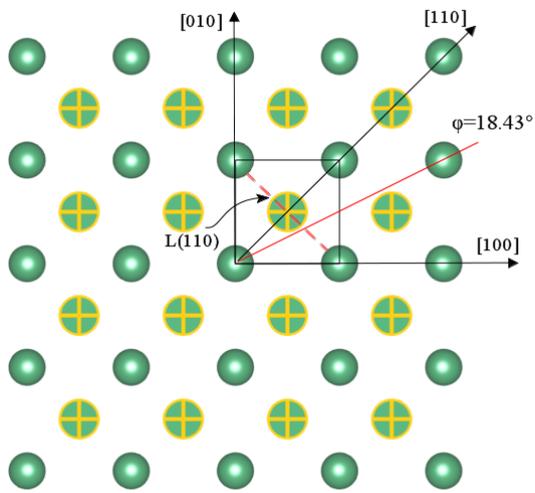

(b)

Figure 4 The projections of atoms in different planes of bcc crystal. In the figure, solid balls are atoms in the plane, and balls with ⊕ are the projections of atoms in nearby planes. (a) ($\bar{1}10$) plane; (b) (001) plane.

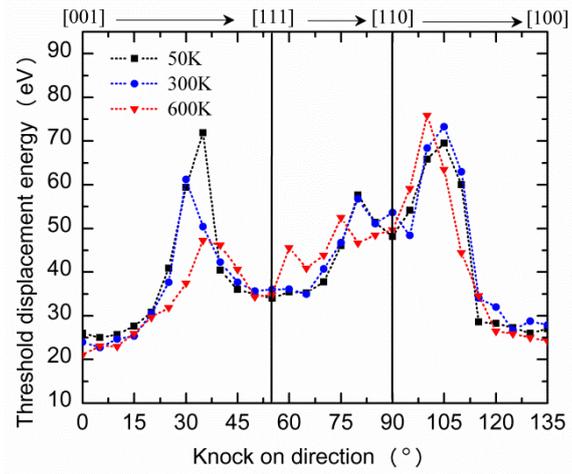

Figure 4. Dependence of the threshold displacement energy calculated by present potential for Nb along different PKA knock-on directions at different temperatures.

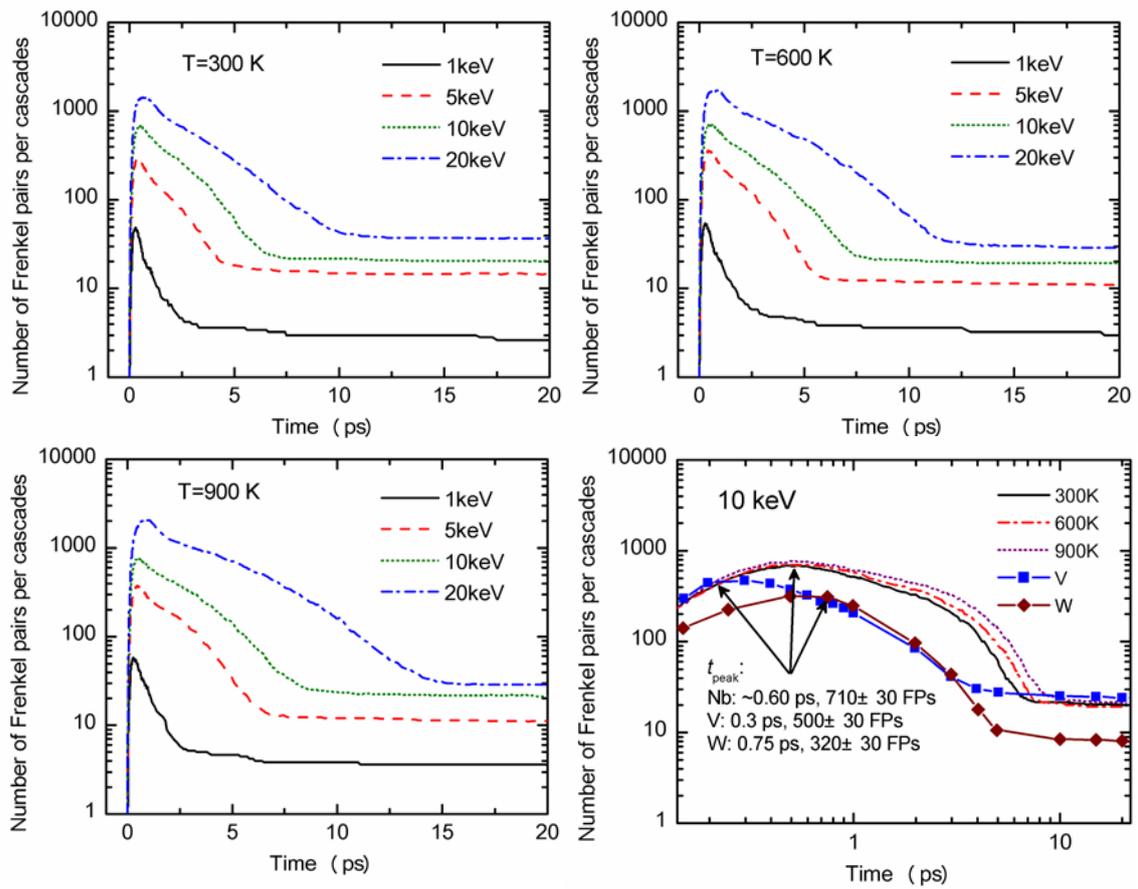

Figure 5. The number of Frenkel pairs as a function of simulated time calculated at different temperatures, and comparisons with Björkas et al.'s [19] results for V and W during 10 keV cascades at 300 K.

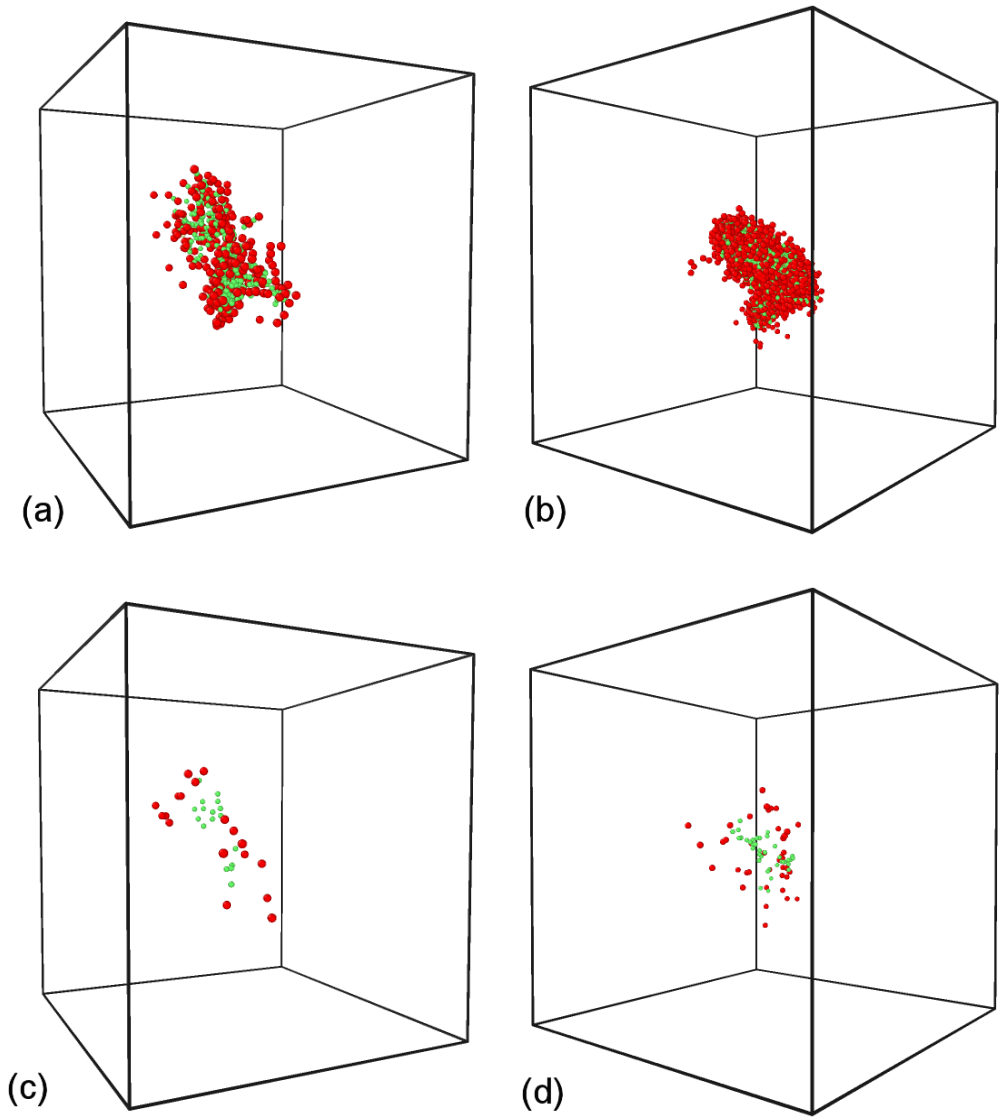

Figure 6. Typical cascade structures of Nb initiated by PKA of 5 keV (a,c) and 20 keV (b,d) energy at 900 K. Peak damage states are shown in (a,b), and final states are shown in (c,d). In the figure, big red solids are interstitials, and small light green solids are vacancies. The size of compute cells are 40×42×44 and 84×86×88 for 5 keV and 20 keV simulations, respectively.

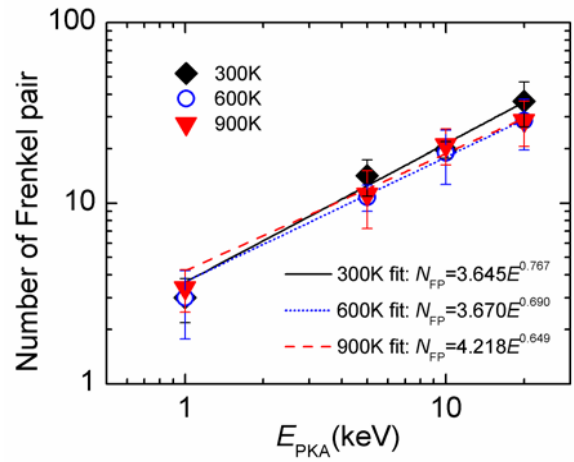

Figure 7 Average number of surviving Frenkel pairs versus PKA energy, and the lines are the power law fit of the calculated data.

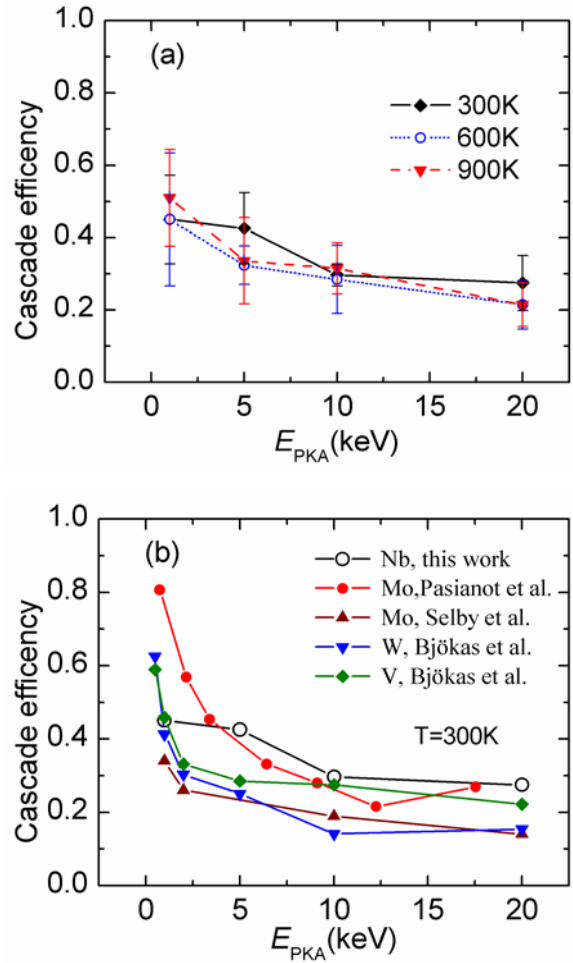

Figure 8 Cascade efficiency versus PKA energy obtained in this work for Nb (a) and its comparisons with results for other refractory metals at 300 K (b).

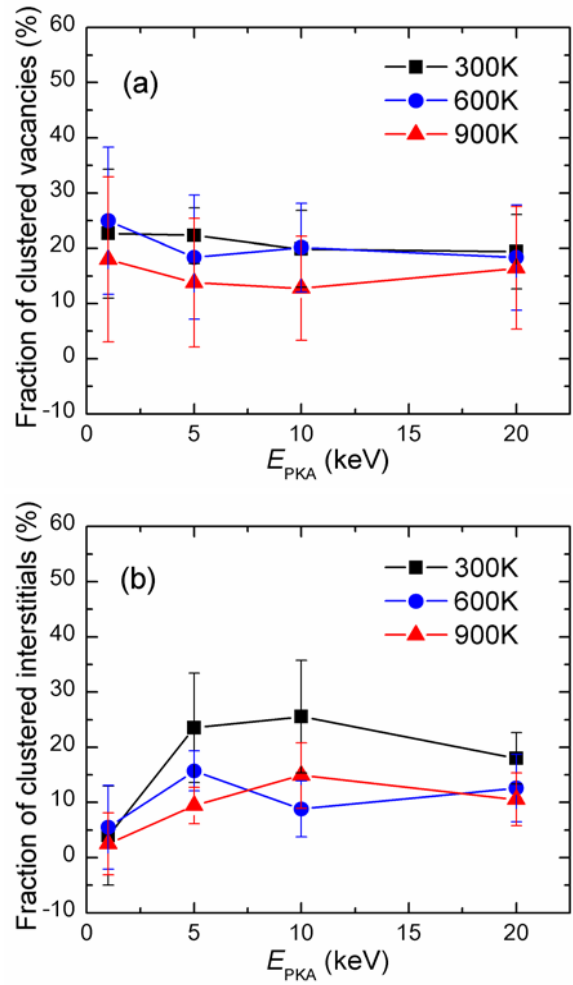

Figure 9 Fraction of clustered defects versus PKA energy.
(a) Vacancy; (b) Interstitial.

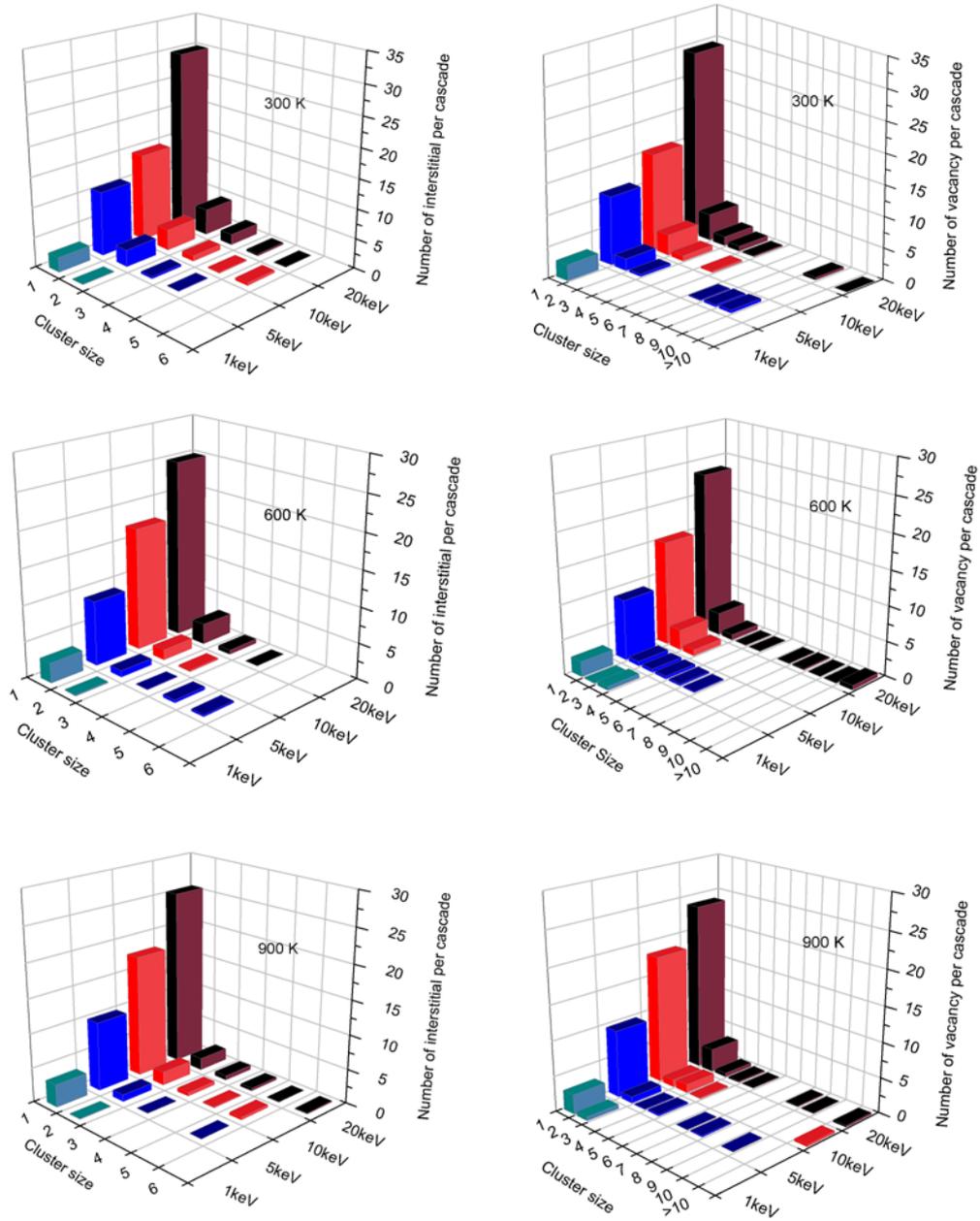

Figure 10 The number of interstitials (left) and vacancies (right) in clusters per cascade as a function of PKA energy.